\documentclass[prd,preprintnumbers,superscriptaddress,nofootinbib,showpacs,twocolumn]{revtex4-2}
\usepackage{pstricks}
\usepackage{color}
\usepackage{amssymb,amsmath,bm,bbm}
\usepackage{epsf,epsfig}
\usepackage{afterpage}
\usepackage{longtable}
\usepackage{latexsym,mathrsfs,dsfont}
\usepackage{graphics}
\usepackage{url}
\usepackage{paralist}
\usepackage{bbold}
\usepackage{appendix}
\usepackage{tikz}
\usepackage[colorlinks=true, urlcolor=blue]{hyperref}

\newcommand{\spur}[1]{\not\! #1 \,}

\newcommand{\be}{\begin{equation}}
\newcommand{\ee}{\end{equation}}
\newcommand{\bi}{\begin{itemize}}
\newcommand{\ei}{\end{itemize}}
\newcommand{\ba}{\begin{array}}
\newcommand{\ea}{\end{array}}
\newcommand{\bea}{\begin{eqnarray}}
\newcommand{\eea}{\end{eqnarray}}
\newcommand{\bec}{\begin{center}}
\newcommand{\eec}{\end{center}}

\newcommand{\nn}{\nonumber}

\def\@seccntformat#1{\@ifundefined{#1@cntformat}%
   {\csname the#1\endcsname\quad}%      default
   {\csname #1@cntformat\endcsname}%    enable individual control
}

\begin{document}

\preprint{BARI-TH/770-25}

\title{   
The charming case of $X(3872)$ and $\chi_{c1}(2P)$ }
\author{Pietro~Colangelo}
\email[Electronic address:]{pietro.colangelo@ba.infn.it} 
\affiliation{Istituto Nazionale di Fisica Nucleare, Sezione di Bari, via Orabona 4, 70126 Bari, Italy}
\author{Fulvia~De~Fazio}
\email[Electronic address:]{fulvia.defazio@ba.infn.it} 
\affiliation{Istituto Nazionale di Fisica Nucleare, Sezione di Bari, via Orabona 4, 70126 Bari, Italy}
\author{Giuseppe~Roselli}
\email[Electronic address:]{g.roselli10@studenti.uniba.it} 
\affiliation{Dipartimento Interateneo di Fisica "Michelangelo Merlin", Universit\`a degli Studi di Bari, via Orabona 4, 70126 Bari, Italy}

\begin{abstract}
\noindent 
More than twenty years have elapsed since the discovery of $\chi_{c1}(3872)$ previously denoted as $X(3872)$, and  an impressive amount of
theoretical and experimental  studies has  been devoted to  its properties, decays and production
mechanisms. Despite  the extensive work, a full understanding of the  nature of $\chi_{c1}(3872)$
is still missing. 
In  the present study we reconsider  a theoretical framework based on the heavy quark large mass limit to
analyze the radiative decays of heavy quarkonia,  in particular  the electric dipole transitions of  $\chi_{c1}(2P)$  to $S$-wave charmonia.  The results favorably compare to recent measurements for  $\chi_{c1}(3872)$,  obtained by the LHCb Collaboration,   if this meson  is identified with $\chi_{c1}(2P)$.  

\end{abstract}

\thispagestyle{empty}

%\vspace*{2cm}
%\pacs{12.60.-i , 13.25.Hw}

\maketitle
\section{Introduction}\label{intro}
The 21st century has witnessed the emergence of a new hadron spectroscopy, in coincidence with a huge number of newly observed states considered as {\it unconventional}, meaning that their properties do not allow to unambiguously identify them as quark-antiquark mesons or three quark baryons. 
This has prompted a number of {\it exotic} interpretations for their structure. These include     loosely bound  hadronic molecules,  compact multiquark states, i.e. tetraquarks and pentaquarks, hybrids having valence quarks plus  gluonic degrees of freedom \cite{Meyer:2015eta}, as well as the possibility that the observed states do not correspond to resonances but to kinematical singularities in the mass distributions deriving, for example, from two body thresholds or triangle singularities \cite{Guo:2019twa}.

Understanding the way quarks bind together to form hadrons would provide us a deeper insight into the nonperturbative regime of quantum chromodynamics, which makes this investigation one of the most fundamental for particle physics.

Most of the observed problematic states contain heavy quarks,  the majority  two heavy quarks.
In the case of mesons this implies that  the exotic interpretation competes with  that of ordinary quarkonium states. For these it is customary to adopt the  spectroscopic classification in terms of quantum numbers: $n^{2s+1}L_J$ where $n$ is the radial quantum number, $L$ and $J$ the orbital and total angular momentum, respectively, $s$ is the spin of the two quarks. A meson with such quantum numbers has parity $P=(-1)^{L+1}$ and charge conjugation $C=(-1)^{L+s}$. 

This paper is mostly devoted to the first, most famous of these quarkoniumlike exotic candidates: the state initially denoted as $X(3872)$, now named $\chi_{c1}(3872)$ \cite{ParticleDataGroup:2024cfk}. It has been first observed by Belle Collaboration \cite{Belle:2003nnu} in the decay $B \to K \pi^+ \pi^- J/\psi$ as a narrow peak in the $\pi^+ \pi^- J/\psi$  invariant mass distribution. Since then, other experiments have confirmed the observation \cite{CDF:2003cab,BaBar:2004iez,D0:2004zmu,LHCb:2011zzp,BESIII:2013fnz,CMS:2013fpt,ATLAS:2016kwu}.  Its quantum numbers have been fixed to $J^{PC}=1^{++}$ \cite{LHCb:2013kgk,LHCb:2014jvf}. Moreover, isospin partners have not been found allowing to establish also $I^G=0^+$ \cite{ParticleDataGroup:2024cfk}. Its narrow width has been determined by means of studies of the line shape \cite{LHCb:2020xds,Belle:2023zxm}.

On the basis of the quantum numbers, if one is convinced that $\chi_{c1}(3872)$ is an   ordinary charmonium state, the most suitable identification is with  $\chi_{c1}(2P)$.
However,
a few puzzling features have prompted different interpretations.

The Particle Data Group (PDG) reports the average value for the mass $m({\chi_{c1}(3872)})=3871.64 \pm 0.06$ MeV, which is remarkably close to the $D^0{\overline D}^{*0}$ threshold $m({D^0)}+m({\overline D}^{*0})=3871.69 \pm 0.07$ MeV. This feature has suggested  that it could be a loosely bound $D^0 {\overline D}^{*0}+{\overline D}^0 D^{*0}$ molecule \cite{Braaten:2003he,Swanson:2003tb,Wong:2003xk,Hanhart:2007yq}. 
The molecular interpretation seems to be contradicted by the small production cross section 
\cite{Artoisenet:2009wk,Bignamini:2009sk}, compatible instead with the  expectations for the ordinary charmonium 
\cite{CDF:2003cab,LHCb:2021ten}.
Other interpretations have also been put forward. They include the compact tetraquark configuration, the mixing between a conventional charmonium and a molecule, the state dynamically generated by a coupled channel effect (for a recent review see \cite{Brambilla:2019esw}).

Another intriguing experimental fact is the measured ratio $\displaystyle\frac{{\cal B}(\chi_{c1}(3872) \to J/\psi \pi^+ \pi^- \pi^0)}{{\cal B}(\chi_{c1}(3872) \to J/\psi \pi^+ \pi^-)}$, experimentally close to 1 \cite{Belle:2005lfc,BaBar:2010wfc,BESIII:2019qvy}. Considering that the two and three  pion states come from $\rho^0$ and $\omega$ decays, respectively, one is led to conclude that isospin is badly violated if $\chi_{c1}(3872)$ has $I=0$ as for the ordinary charmonium. However, the three pion state is kinematically strongly suppressed due to phase space limitations \cite{Suzuki:2005ha}, potentially mitigating this argument. A recent LHCb analysis \cite{LHCb:2022jez} shows that, even taking into account this observation,  the isospin violation is larger than  in  $\psi(2S)$ decays \cite{ParticleDataGroup:2024cfk}. Therefore, this issue remains an open puzzle.

\begin{table}[t]
\centering 
\begin{tabular}{l c} 
\hline \hline
Collaboration & ${\cal R}(\chi_{c1}(3872))$   \\ 
\hline
BABAR 2008 \cite{BaBar:2008flx} & $3.4 \pm 1.4$ \\
Belle 2011 \cite{Belle:2011wdj} & $<2.1 \, (90\% \, {\rm C.L.})$ \\ 
LHCb 2014 \cite{LHCb:2014jvf} & $2.46 \pm 0.64 \pm 0.29$ \\ 
BESIII 2020 \cite{BESIII:2020nbj} & $<0.59 \, (90 \% \, {\rm C.L.}) $ \\ 
LHCb 2024 (run 1) & $ 2.50 \pm 0.52 ^{+0.20}_{-0.23} \pm0.06 $ \\ 
LHCb 2024 (run 2) & $ 1.49 \pm 0.23 ^{+0.13}_{-0.12} \pm0.03 $ \\
LHCb 2024 average \cite{LHCb:2024tpv} & \, $ 1.67 \pm 0.21 \pm 0.12 \pm0.04 $ \, \\ \hline \hline
\end{tabular} 
\caption{Experimental results for ${\cal R}(\chi_{c1}(3872))$. } 
\label{tabRexp} 
\end{table} 
%%%%%%%%%%%%%%%%%%%%%

It has been put forward that an observable  sensitive to the structure of $\chi_{c1}(3872)$ is the ratio  ${\cal R}(\chi_{c1}(3872))=\displaystyle\frac{{\cal B}(\chi_{c1}(3872) \to \psi(2S) \gamma)}{{\cal B}(\chi_{c1}(3872) \to J/\psi \gamma )}$ \cite{Swanson:2004pp}. While several studies relying on the ordinary charmonium interpretation predict  ${\cal R}\ge 1$, in the case of exotic configurations smaller values are obtained. These results have to be contrasted with the  most recent experimental measurement from LHCb Collaboration providing ${\cal R}=1.67 \pm 0.21 \pm 0.12 \pm 0.04$ \cite{LHCb:2024tpv}, which corresponds to the average of the 2024 results of run 1 and run 2 quoted in  Table \ref{tabRexp}. Previous experimental results are also collected in the table.

In order to probe the identification of $\chi_{c1}(3872)$ with $\chi_{c1}(2P)$, in this paper we  work out the phenomenological consequences of such an  interpretation,  deriving predictions for experimentally measured/measurable properties of $\chi_{c1}(2P)$ once it is assigned the mass $M=3872$ MeV and the width $\Gamma=1.19 \pm 0.21$ MeV \cite{ParticleDataGroup:2024cfk}.
At the beginning of our study we stress that, while it is possible that a chosen  interpretation  leads to verified predictions despite of being wrong, the definitive disproval of the identification $X(3872)=\chi_{c1}(2P)$ would be the discovery of another particle with mass in the same range of $\chi_{c1}(3872)$ and properties  indicating that it is a more suitable candidate to be identified with $\chi_{c1}(2P)$. No exotic interpretation could be validated if the ordinary charmonium spectrum is not filled with the $\chi_{c1}(2P)$ state.

\section{Heavy quark symmetries at work}
In this section we adopt an effective Lagrangian approach exploiting the heavy quark spin symmetry to study the properties and decays of heavy quarkonia, i.e. the states $Q{\bar Q}$.

The physics of hadrons containing  one  heavy quark (HQ) can be systematically studied considering the 
 large heavy quark mass limit, which is formalized in the heavy quark effective theory (HQET)  \cite{Neubert:1993mb}.
In this limit the HQ can be treated as a static color source and is almost decoupled from the rest of the hadron  resulting 
in invariance of strong interactions under HQ spin and/or flavor rotations.
To derive the HQET Lagrangian one introduces the field 
$h_v(x)=e^{im_Q v \cdot x}P_+Q(x)$ where $Q$ is the HQ field in QCD and $v$ is the HQ velocity. The field 
 $h_v$ satisfies 
$h_v=P_+ Q$ where $P_+=\displaystyle{1 + \spur v \over 2}$
so that the relation holds:
${\spur v}h_v=h_v$.
The HQ is off shell and its momentum can be written as $p=m_Q v+k$ where $k$ is a residual momentum of ${\cal O}(\Lambda_{QCD})$.
In terms of $h_v$ the HQET Lagrangian reads ${\cal L}_{HQET}={\bar h}_v i \, v \cdot D h_v$, $D$ being the QCD covariant derivative. For $N_f$ heavy flavors this is invariant under $ SU(N_f)$ spin/flavor rotations. Symmetry breaking terms arise including subleading operators suppressed by powers of $k/m_Q$. At ${\cal O}(1/m_Q)$ two operators appear
\be
{\cal L}^{(1)}={1 \over 2 m_Q} {\bar h}_v (i \spur
D_\perp)^2 h_v +{1 \over 2 m_Q}{\bar h}_v {g_s \sigma_{\alpha \beta}
G^{\alpha \beta} \over 2 } h_v 
\label{lag1m} \; ,
\ee
they represent the  kinetic energy of the heavy quark
due to its residual momentum $k$, and the
chromomagnetic coupling of the heavy quark spin to the gluon field, respectively.

Heavy quark symmetries have several consequences of great relevance for phenomenology of systems with a single heavy quark. Spectroscopic implications for such systems have been first pointed out in \cite{Isgur:1991wq}.

Here instead we are interested in systems with two heavy quarks, in particular mesons, i.e. quarkonia consisting in a heavy quark-antiquark pair. 
For these systems flavor symmetry cannot be exploited 
 \cite{Thacker:1990bm}.
The reason is that, when considering gluon exchanges between two heavy quarks having the same velocity, infrared divergences show up. They can be regulated  going beyond the leading order in the heavy quark expansion of the QCD Lagrangian, specifically including  the kinetic energy operator that  breaks flavor symmetry. 
On the other hand, spin symmetry holds  both for heavy-light systems and  for systems with two heavy quarks.

Due to such a  symmetry, hadrons which differ only for the orientation of the heavy quark spin can be collected in multiplets, which would be  degenerate in the infinite HQ mass limit. The degeneracy is  broken at first by terms of ${\cal O}(1/m_Q)$, in particular by the chromomagnetic operator.

Considering mesons, in the case of heavy-light configurations each  multiplet comprises two states. For such systems, exploiting the HQ symmetries, the application of   an effective Lagrangian approach to compute the strong decays to light pseudoscalar and vector mesons has provided a number of successful predictions \cite{Colangelo:2003vg,Colangelo:2004vu,Colangelo:2006rq,Colangelo:2007ds,Colangelo:2010te,Colangelo:2012xi,Campanella:2018xev}.

For quarkonia it is possible to rotate not only the spin of one HQ but also the spin of the heavy antiquark. As a result, one can organize the heavy quarkonia in multiplets populated by a number of states which depends on the orbital angular momentum.
A multiplet of $Q{\bar Q}$ states  with relative orbital angular momentum $L$ is described by
\begin{widetext}
\bea  J^{\mu_1 \dots \mu_L}&=&{ 1+ \spur{v} \over 2}\Big(
H_{L+1}^{\mu_1 \dots \mu_L \alpha } \gamma_\alpha + {1 \over
\sqrt{L(L+1)}} \sum_{i=1}^{L} \epsilon^{\mu_i \alpha \beta \gamma}
v_\alpha \gamma_\beta H_{L \gamma}^{\mu_1 \dots \mu_{i-1}
\mu_{i+1} \dots \mu_L}\nonumber \\ &+& {1 \over L} \sqrt{2L-1
\over 2L+1} \sum_{i=1}^{L}(\gamma^{\mu_i} -v^{\mu_i})
H_{L-1}^{\mu_1 \dots \mu_{i-1} \mu_{i+1} \dots
\mu_L}\label{multiplet} \\& -&{2 \over L \sqrt{(2L-1)(2L+1)}}
\sum_{i<j} (g^{\mu_i \mu_j}-v^{\mu_i}v^{\mu_j}) \gamma_\alpha
H_{L-1}^{\alpha \mu_1 \dots \mu_{i-1} \mu_{i+1} \dots \mu_{j-1}
\mu_{j+1}\dots \mu_L} \nonumber \\ &+&K_L^{\mu_1 \dots \mu_L}
\gamma_5 \Big){ 1- \spur{v} \over 2} \,\,\, , \nonumber \eea
\end{widetext}
\noindent where  $v^\mu$ is  the heavy quark four-velocity and the transversality condition $v_{\mu_i}J^{\mu_1 \dots \mu_L}=0$, $i=1 \dots L$ holds.
$H_A$, $K_A$ are the effective fields of the various members of
the multiplets with total spin $J=A$. Since  we  are mainly interested in $S-$ and $P-$wave states, i.e. those  corresponding to $L=0,\,1,$ as in Tables \ref{ccbar} and \ref{bbbar},   we list 
the  multiplets   obtained from (\ref{multiplet}),
\begin{itemize}
\item L=1 multiplet: \bea  J^{\mu }&=&{ 1+ \spur{v} \over 2}\Big\{
H_2^{\mu \alpha } \gamma_\alpha  + {1 \over \sqrt{2}}
\epsilon^{\mu \alpha \beta \gamma} v_\alpha \gamma_\beta H_{1
\gamma}\nonumber
\\&+& {1 \over \sqrt{3}}
(\gamma^{\mu} -v^{\mu}) H_0  + K_1^{\mu }\gamma_5 \Big\}{ 1-
\spur{v} \over 2}\,\,; \label{Pwave} \eea \item L=0 multiplet: \be
J={ 1+ \spur{v} \over 2} \left[H_1^\mu \gamma_\mu -H_0 \gamma_5
\right]{ 1- \spur{v} \over 2} \,\,.\label{Swave} \ee
\end{itemize}
In the following we consider the application of the HQ symmetries  to radiative decays of quarkonia. 
The purposes are manifold.
Exploiting the available experimental data  we derive a number of universal parameters entering in the description of such decays in the HQ limit. 
At the same time, we check the reliability of the HQ spin symmetry.
Considering radiative decays of  $\chi_{c1}(2P)$, assigning to this particle the properties of $\chi_{c1}(3872)$, i.e. the mass and full width,   we discuss whether experimental data can support the identification $\chi_{c1}(3872)=\chi_{c1}(2P)$.
Several predictions for modes not yet observed are also provided, they would serve as a further test of the approach once more data will be available.
Moreover, taking into account the first symmetry breaking terms responsible for the mass splittings among the members of the multiplets, we compute the parameters  determining such terms,  finding relations among them.

 Let us conclude this section with two remarks. First,  the multiplet in Eq.~\eqref{multiplet}, as well as the analogous multiplets constructed for other values of the orbital angular momentum $L$,  is constructed applying the heavy quark symmetries both to $Q$ and $\bar Q$.  One might wonder if this expression could also describe a cryptoexotic meson, i.e. a tetraquark with non exotic quantum numbers.  Since the multiplet \eqref{multiplet} is  an $SU(3)$ singlet, a case that could be  considered would be a tetraquark with light degrees of freedom in  isoscalar scalar configuration, decoupled from the heavy quark complex. However, the effective description of multiquark mesons comprising pairs of  heavy and light quarks deals with a difficult multi scale problem,  requiring dedicated studies. As for the electromagnetic processes, in general the e.m. interaction Lagrangian, discussed in the next section,  should also be extended to describe to photon coupled to the  light degrees of freedom, which is a long-distance process. 

The second remark concerns a comparison of the approach we have followed and nonrelativistic QCD (NRQCD). A difference is that  the effective theory we are using, see the currents  \eqref{multiplet}, involves   hadrons appearing  in effective interaction Lagrangians constructed using the heavy quark symmetries. On the other hand, in NRQCD one works out effective operators obtained from the heavy quark expansion of the QCD Lagrangian, which must be considered in matrix elements of external hadron states.

\section{Radiative decays of heavy quarkonia}
%%%%%%%%%%%%%%%%%%%%%%%%%%%%%%%%%%%%%%%%%%%%%%
\begin{table}[t]
\centering
\begin{tabular}{l c c  c}
\hline \hline
%\rowcolor{red!20}
State & J$^{PC}$ & Mass (MeV) & $\Gamma_{tot}$  (MeV) \\
\hline
$\eta_c(1S)$ & $0^{-+}$ & $2984.1 \pm 0.4$ & $30.5 \pm 0.5$  \\
$J/\psi$ & $1^{--}$ & $3096.900 \pm 0.006$ & \, $(92.6 \pm 1.7 )\times 10^{-3}$ \, \\
$\chi_{c0}(1P)$ & $0^{++}$ & $3414.71 \pm 0.30$ & $10.5 \pm 0.8$ \\
$\chi_{c1}(1P)$ & $1^{++}$ & $3510.67 \pm 0.05$ & $0.88 \pm 0.05$  \\
$\chi_{c2}(1P)$ & $2^{++}$ & $3556.17 \pm 0.07$ & $2.00 \pm 0.11$  \\
$h_c(1P)$ & $1^{+-}$ & $3525.37 \pm 0.14$ & $0.78 \pm^{0.27}_{0.24 } \pm 0.12$  \\
$\eta_c(2S)$ & $0^{-+}$ & $3637.7 \pm 0.9$ & $11.8 \pm 1.6$  \\
$\psi(2S)$ & $1^{--}$ & $3686.097 \pm 0.010$ & $(286 \pm 16) \times 10^{-3}$  \\
$\chi_{c0}(3860)$ & $0^{++}$ & $3862 ^{+26+40}_{-32-13}  \,$ &  $201 ^{+154+88}_{-67-82} $ \\
$\chi_{c1}(3872)$ & $1^{++}$ & $3871.64 \pm 0.06$ & $1.19 \pm 0.21$  \\
$\chi_{c2}(3930)$ & $2^{++}$ & $3922.5 \pm 1.0$ & $35.2 \pm 2.2$  \\
\hline \hline
\end{tabular}
\caption{Properties of $c\bar{c}$ mesons \cite{ParticleDataGroup:2024cfk}. In this work we identify the last three states with those in the 2P multiplet. }
\label{ccbar}
\end{table}
%%%%%%%%%%%%%%%%%%%%%%%%%%%%%%%%%%%%%%%%%%%%%%%%

\begin{table}[b]
\centering
\begin{tabular}{l c c c }
\hline \hline
%\rowcolor{blue!20}
State & J$^{PC}$ & Mass (MeV) & $\Gamma_{tot}$  (MeV) \\
\hline
$\eta_b(1S)$ & $0^{-+}$ & $9398.7 \pm 2.0$ & $10 ^{+5}_{-4}$ \\
$\Upsilon(1S)$ & $1^{--}$ & $9460.40 \pm 0.09 \pm 0.04$ & $(54.02\pm 1.25) \times 10^{-3}$ \\
$\chi_{b0}(1P)$ & $0^{++}$ & $9859.44 \pm 0.42 \pm 0.31$ &  \\
$\chi_{b1}(1P)$ & $1^{++}$ & $9892.78 \pm 0.26 \pm 0.31$ & \\
$\chi_{b2}(1P)$ & $2^{++}$ & $9912.21 \pm 0.26 \pm 0.31$ & \\
$h_b(1P)$ & $1^{+-}$ & $9899.3 \pm 0.8$ & \\
$\eta_b(2S)$ & $0^{-+}$ & $9999.0 \pm 3.5 {}_{-1.9}^{+2.8}$ & $<24$  \\
$\Upsilon(2S)$ & $1^{--}$ & $10023.4 \pm 0.5$ & $(31.98 \pm 2.63) \times 10^{-3}$  \\
$\chi_{b0}(2P)$ & $0^{++}$ & $10232.5 \pm 0.4 \pm 0.5$ & \\
$\chi_{b1}(2P)$ & $1^{++}$ & $10255.46 \pm 0.22 \pm 0.50$ &  \\
$\chi_{b2}(2P)$ & $2^{++}$ & $10268.65 \pm 0.22 \pm 0.50$ &  \\
$h_b(2P)$ & $1^{+-}$ & $10259.8 \pm 0.5 \pm 1.1$ &  \\
\hline \hline
\end{tabular}
\caption{Properties of $b\bar{b}$ mesons \cite{ParticleDataGroup:2024cfk}.}
\label{bbbar}
\end{table}
%%%%%%%%%%%%%%%%%%%%%%%%%%%%%%%%%%%%%%%%%%%%%%%%

Let us start  describing the
radiative decays among members of the $P-$wave  and of the
$S-$wave multiplets. Since in such processes the orbital angular momentum of the decaying and produced quarkonia differ for $\Delta L=1$, these are electric dipole transitions. The effective Lagrangian governing the decays has been derived in \cite{Casalbuoni:1992yd} in the approximation of assuming  that quarks 
exchange soft gluons. It reads
\be {\cal L}_{nP \leftrightarrow mS}=\delta^{nPmS}_Q {\rm Tr}
\left[{\bar J}(mS) J_\mu(nP) \right] v_\nu F^{\mu \nu} + \rm{h.c.}
\,.\label{lagPS} \ee 
 $F^{\mu \nu}$ is the electromagnetic field
strength tensor and $\delta^{nPmS}_Q$ is a constant. It can be checked that  the Lagrangian \eqref{lagPS} is invariant under the discrete parity $P$,  charge conjugation $ C$ and time reversal $T$ transformations
\bea
 J^{\mu_1 \dots \mu_L}&&
\stackrel{P}{\rightarrow}\gamma^0 J_{\mu_1 \dots \mu_L} \gamma^0
\label{parity} \\
J^{\mu_1 \dots \mu_L}&&
\stackrel{C}{\rightarrow}(-1)^{L+1} C [J^{\mu_1 \dots \mu_L}]^T C
\label{charge-conj} \\
 J^{\mu_1 \dots \mu_L}&&
\stackrel{T}{\rightarrow}-T J_{\mu_1 \dots \mu_L} T^{-1} \,\, ,
\label{time-reversal} \eea
where $C=i \gamma^2 \gamma^0$ and $T=i \gamma^1 \gamma^3$.
Moreover, the Lagrangian in \eqref{lagPS} is invariant under HQ spin transformations
\be J^{\mu_1 \dots \mu_L}
\stackrel{SU(2)_{S_h}}{\rightarrow}S J^{\mu_1 \dots \mu_L}
S^{\prime \dagger} \,\,.\label{HQspin} \ee 
In (\ref{HQspin}) we have $S$ and 
$S^\prime\in SU(2)_{S_h}$,  $SU(2)_{S_h}$  being the group of heavy
quark spin rotations, with the relations $[S,{\spur
v}]=[S^\prime,{\spur v}]=0$.

As a consequence of spin symmetry, the constant $\delta^{nPmS}_Q$  in \eqref{lagPS} 
describes all transitions among the members of the $nP$
multiplet and those of the $mS$ one. 
In terms of this parameter 
 the following decay
widths are obtained from (\ref{lagPS}) \cite{Casalbuoni:1992yd,DeFazio:2008xq},
\bea
\Gamma(n^3P_J \to m^3S_1 \gamma) &=& {(\delta_Q^{nPmS})^2 \over
3 \pi} k_\gamma^3{M_{S_1} \over M_{P_J}}   \label{deltanPmS1} \\
\Gamma(m^3S_1 \to n^3P_J \gamma) &=& (2J+1) {(\delta_Q^{nPmS})^2
\over 9 \pi} k_\gamma^3{M_{P_J} \over M_{S_1}}\nn \\  \label{deltanPmS2} \\
\Gamma(n^1P_1 \to m^1S_0 \gamma) &=& {(\delta_Q^{nPmS})^2 \over 3
\pi} k_\gamma^3{M_{S_0} \over M_{P_1}} \label{deltanPmS3}\\
\Gamma(m^1S_0 \to n^1P_1 \gamma) &=& {(\delta_Q^{nPmS})^2 \over
\pi} k_\gamma^3{M_{P_1} \over M_{S_0}}  \label{deltanPmS4}
 \,,\eea where $k_\gamma$ denotes
the photon energy.

An analogous Lagrangian can be written for the decays between the doublets $nS \to mS$ with $\Delta L=0$, the  magnetic dipole transitions
\be
 {\cal L}_{nS \leftrightarrow mS}=\delta^{nSmS}_Q {\rm Tr}
\left[{\bar J}(mS) \sigma_{\mu \nu}J(nS) \right]  F^{\mu \nu} + \rm{h.c.}
\,.\label{lagSS} \ee 
The widths of the corresponding radiative transitions are given by
\be
\Gamma(n^3S_1 \to m^1S_0 \gamma) = {4(\delta_Q^{nSmS})^2 \over
3 \pi} k_\gamma^3{M_{S_0} \over M_{S_1}}   \,\,.\label{deltanSmS} \\
\ee
The Lagrangian \eqref{lagSS} is also  invariant under $P,\,C$ and $T$. However, it violates the spin symmetry because of the presence of $\sigma_{\mu \nu}$ \cite{Cho:1992nt}. Nevertheless, interesting information can be obtained from it.

\section{Mass splittings within the multiplets}
Degeneracy among the members of a given spin multiplet is broken by the action of the chromomagnetic operator in \eqref{lag1m}. 
In the effective Lagrangian approach this can be represented by the  expression
\be
{\cal L}^{SpSB}=
\frac{\lambda_J }{2m_Q}\,\frac{1}{\cal N}{\rm Tr} [{\bar J}_{\mu_1 \dots \mu_n}\sigma_{\alpha \beta} J_{\nu_1 \dots \nu_n}\Phi^{\mu_1 \dots \mu_n \alpha \beta \nu_1 \dots \nu_n}]
\label{splitting-gen}
\ee
where the superscript stays for spin symmetry breaking, and ${\cal N}={\rm Tr} [{\bar J}_{\mu_1 \dots \mu_n} J^{\mu_1 \dots \mu_n}]$ takes into account the normalization of the states.
The structure of the function $\Phi$ is  such that the previous term is invariant under the discrete symmetries $P,\,C,\,T$; the antisymmetry of $\sigma_{\alpha \beta}$ must also be taken into account. Finally, none of the indices of $\Phi$ can be carried by  the velocity $v$ in view of the transversality condition. 
In the case of $S-$wave states, the only expression is
\be
{\cal L}_S^{SpSB}=\frac{\lambda_S}{2m_Q} \,\frac{1}{\cal N}{\rm Tr}  [{\bar J}\sigma_{\alpha \beta} J\,\sigma^{ \alpha \beta }]
\label{splitting-S}
\ee
that represents the hyperfine splitting due to spin-spin interaction.
The   $S-$wave doublet  consists of two mesons with spin, parity and charge conjugation  $J^{PC}=(0^{-+},\,1^{--})$; their masses obey the expansion [at order ${\cal O}(1/m_Q)]$,
\be
m_H=m_Q +{\bar \Lambda}_S+ \frac{-\lambda_1+d_H \, \lambda_S}{2m_Q}\,\,. \label{massformula}
\ee
 The parameter $\lambda_1$ is related to the heavy quark kinetic energy and does not break spin symmetry. The spin splitting is instead due to the factor  $d_H$ that multiplies
$\lambda_S$,  which is different for the two states in the doublet. It can be fixed  using the trace formalism in Eq.~\eqref{splitting-S},  so that  it is possible to relate $\lambda_S$ to the observed mass splitting,
\be
\lambda_S=\frac{1}{8}\left(m_V^2-m_P^2 \right) \,\,.
\ee
In the case of $P$-wave states one can write
\be
{\cal L}_P^{SpSB}=
\frac{\lambda_P}{2m_Q} \,\frac{1}{\cal N}{\rm Tr} [{\bar J}_{\mu}\sigma_{\alpha \beta} J_{\nu}\Phi^{ \alpha \beta  \mu \nu}]\,\,.
\label{splitting-P}
\ee
The function $\Phi^{ \alpha \beta \mu \nu}$ satisfying the previous requirements can be written as
\bea
\lambda_P \Phi^{ \alpha \beta \mu \nu}&=&\lambda_P^{(1)}\, {\rm Tr}[{\bar J}_{\mu}\sigma_{\alpha \beta} J_{\mu}\sigma^{ \alpha \beta}] \label{DmP} \\
&&\hskip-0.7cm +i \,\lambda_P^{(2)}\, \left({\rm Tr}[{\bar J}^{\alpha}\sigma_{\alpha \beta} J^{\beta}]+{\rm Tr}[ J^{\alpha}\sigma_{\alpha \beta} {\bar J}^{\beta}]  \right) \nn \\
&&\hskip-0.7cm +\lambda_P^{(3)}\, \left({\rm Tr}[{\bar J}_{\mu}\sigma_{\alpha \beta} J^{\beta}\sigma^{\alpha \mu}]+{\rm Tr}[ J_{\mu}\sigma^{\alpha \mu} {\bar J}^{\beta}\sigma_{\alpha \beta}]  \right) \nn \,.
\eea
The $P-$wave multiplet comprises the states with $J^{PC}=(0^{++},\,1^{++},\,2^{++},\,1^{+-})$. 
In analogy to \eqref{massformula} one can write the masses according to
\bea
m_H&=&m_Q +{\bar \Lambda}_P+ \frac{-\lambda_1}{2m_Q}
\nn \\
&+& \frac{d_H^{(1)} \, \lambda_P^{(1)}+d_H^{(2)} \, \lambda_P^{(2)}+d_H^{(3)} \, \lambda_P^{(3)}}{2m_Q} \,\,.
 \label{massformulaP}
\eea
Adopting the procedure described in the previous case one finds
\bea
\lambda_P^{(1)} &=&\frac{1}{24} \left(-m_0^2+3m_1^2+m_2^2-3m_h^2 \right) \nn \\
\lambda_P^{(2)} &=&\frac{1}{24} \left(-2m_0^2-3m_1^2+5m_2^2 \right) \label{lambdaPi} \\
\lambda_P^{(3)} &=&\frac{1}{24} \left(-2m_0^2+3m_1^2-m_2^2 \right) \, , \nn 
\eea
where $m_0,\,m_1,\,m_2$ and $m_h$ denote the masses of the four states in the multiplet with $J^{PC}=0^{++},\,1^{++},\,2^{++},\,1^{+-}$, respectively.

\section{Numerical results: Mass splittings parameters}
We exploit the measured masses of the heavy quarkonia to determine the parameters $\lambda_P^{(i)}$,  with $i=1, 2, 3$, in \eqref{lambdaPi}.
This can be done for the 1P multiplet in the case of charm and for 1P and 2P multiplets in the case of beauty. In the 2P charm multiplet the mass of $h_c(2P)$ is missing, preventing us to apply the procedure as for the other multiplets.
We find
\bea
\lambda_{c,1P}^{(1)}&=& (2.82 \pm 0.02) \times 10^{-2} \,\,{\rm GeV}^2
\nn \\
\lambda_{c,1P}^{(2)}&=&(12.24 \pm 0.02 )\times 10^{-2} \,\,{\rm GeV}^2
\label{c1P} \\
\lambda_{c,1P}^{(3)}&=&(4.20 \pm 0.02) \times 10^{-2} \,\,{\rm GeV}^2
\nn 
\eea
%%%%%%%%%%%%%%%%
\bea
\lambda_{b,1P}^{(1)}&=&(2.75 \pm 0.25)\times 10^{-2}\,\, {\rm GeV}^2
\nn \\
\lambda_{b,1P}^{(2)}&=&(13.5 \pm 0.2 )\times 10^{-2}\,\, {\rm GeV}^2
\label{b1P} \\
\lambda_{b,1P}^{(3)}&=&(3.9 \pm 0.2)\times 10^{-2}\,\, {\rm GeV}^2
\nn 
\eea
%%%%%%%%%%%%%%%%%
\bea
\lambda_{b,2P}^{(1)}&=&(2.0 \pm 0.4) \times 10^{-2}\,\, {\rm GeV}^2
\nn \\
\lambda_{b,2P}^{(2)}&=&(9.6 \pm 0.3 ) \times 10^{-2} \,\,{\rm GeV}^2
\label{b2P} \\
\lambda_{b,2P}^{(3)}&=&(2.8 \pm 0.2 )\times 10^{-2} \,\,{\rm GeV}^2 \,\,.
\nn 
\eea
%%%%%%%%%%%%%%%%%%%

\section{Numerical results: Radiative decays}
The effective Lagrangian approach, based on the HQ limit and exploited to study  the radiative heavy quarkonium decays,  has been used already  in \cite{DeFazio:2008xq}
  predicting ${\cal R}(\chi_{c1}(3872))=1.64 \pm 0.25$, compatible with  the  experimental data available at that time ${\cal R}(\chi_{c1}(3872))=3.4 \pm 1.4$  \cite{BaBar:2008flx} (first entry in Table \ref{tabRexp}). 
In view of the improvement of the experimental data and, in particular, the recent results on the radiative decays of $\chi_{c1}(3872)$, we consider mandatory to refine the theoretical prediction including all the new experimental information.

\begin{table}[t]
\centering 
\begin{tabular}{ l  c } 
\hline  \hline
%\rowcolor{red!20} 
\text{Decay mode} & \text{Branching ratio }  \\
\hline 
$J/\psi \rightarrow\gamma \eta_c(1S)$ & $( 1.41 \pm 0.14 ) \%$  \\
$\chi_{c0}(1P)\rightarrow\gamma J/\psi$ & $(1.41 \pm 0.09) \%$   \\
$\chi_{c1}(1P)\rightarrow\gamma J/\psi$ & $(34.3 \pm 1.3) \%$   \\
$\chi_{c2}(1P)\rightarrow\gamma J/\psi$ & $(19.5 \pm 0.8)\%$  \\
$h_c(1P) \rightarrow\gamma \eta_c(1S)$ & $(60 \pm 4)$ \%   \\
$\eta_c(2S)\rightarrow\gamma J/\psi$ & $<1.4 \% $   \\
$\psi(2S)\rightarrow\gamma \eta_{c}(1S)$   & $(3.6 \pm 0.5) \times 10^{-3}$   \\
$\psi(2S)\rightarrow\gamma \eta_{c}(2S)$   & $(7 \pm 5) \times 10^{-4}$  \\ 
$\psi(2S)\rightarrow\gamma \chi_{c0}(1P)$  & $(9.77 \pm 0.23)\%$  \\
$\psi(2S)\rightarrow\gamma \chi_{c1}(1P)$  & $(9.75 \pm 0.27)\%$   \\
$\psi(2S)\rightarrow\gamma \chi_{c2}(1P)$  & $(9.36 \pm 0.23)\%$  \\
$ \chi_{c1}(3872)\rightarrow\gamma J/\psi $ & \, $ (7.8 \pm 2.9) \times 10^{-3} $  \, \\
\hline  \hline
\end{tabular} 
\caption{Radiative branching ratios of $c\bar{c}$ mesons \cite{ParticleDataGroup:2024cfk}.  }
\label{BRccbar} 
\end{table}
%%%%%%%%%%%%%%%%%%%%%%%%%%%%%%%%%%%%%%%%%%%%%%%%%

\begin{figure}[t]
\begin{center}
\includegraphics[width = 0.5\textwidth]{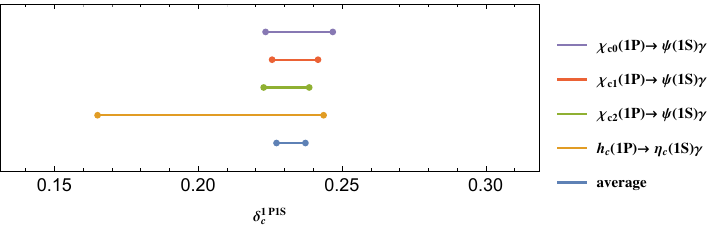}
\caption{\baselineskip 10pt  \small  Coupling $\delta_c^{1P1S}$ obtained comparing the theoretical predictions for the modes listed in the legenda to the available experimental data. The average value is also displayed.}\label{fig:deltac1P1S}
\end{center}
\end{figure}

%%%%%%%%%%%%%%%%%%%%%
We start considering the decays of the charm $P-$wave multiplet with $n=1$, i.e. the states $\chi_{c0}(1P),\,\chi_{c1}(1P),\,\chi_{c2}(1P),\,h_c(1P)$, to the doublet of $S-$wave states with $n=1$: $\eta_c(1S),\,J/\psi$. All these states have been experimentally observed and their masses and total decay widths are measured \cite{ParticleDataGroup:2024cfk}. Possible radiative transitions are $\chi_{cJ} \to J/\psi \gamma$, with  $J=0,1,2$, and $h_c \to \eta_c \gamma$. The transitions $\chi_{cJ} \to \eta_c\gamma$  and $h_c \to J/\psi \gamma$ are forbidden by charge conjugation invariance.
Exploiting the four measured radiative branching fractions (quoted in Table \ref{BRccbar}) it is possible to determine the coupling $\delta^{1P1S}$ comparing them to  Eq.~\eqref{deltanPmS1}, and to  verify if they fulfill the HQ spin symmetry prediction.
We obtain the results shown in Fig.~\ref{fig:deltac1P1S}. In the HQ limit the value of $\delta_c^{nPmS}$  should be the same in all the four cases, a very well satisfied prediction. 
The average value corresponds to
\be
\delta_c^{1P1S}=0.232 \pm 0.05 \,\,{\rm GeV^{-1}}\,\,.
\ee
\begin{figure}[t]
\begin{center}
\includegraphics[width = 0.5\textwidth]{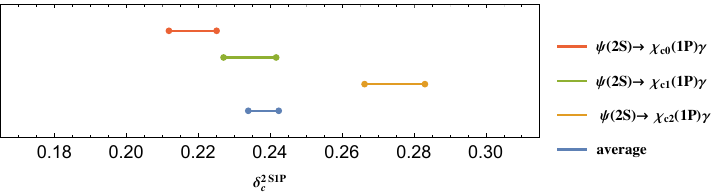}
\caption{\baselineskip 10pt  \small  Coupling $\delta_c^{2S1P}$ obtained comparing the theoretical predictions for the modes listed in the legenda to the available experimental data. The average value is also displayed.}\label{fig:deltac2S1P}
\end{center}
\end{figure}
The same procedure can be carried out for the transitions from the 2S multiplet to the 1P one. Specifically, we consider the transitions $\psi(2S) \to \chi_{cJ}(1P) \gamma$ and $\eta_c(2S) \to h_c(1P) \gamma$. No experimental data are available for  the branching fraction of the latter process, so that we use the first three modes to check the goodness of the HQ spin symmetry in these transitions and to determine the coupling $\delta_c^{2S1P}$. Then, we use the result to predict ${\cal B}(\eta_c(2S) \to h_c(1P) \gamma)$.
Figure~\ref{fig:deltac2S1P} shows that in this case a deviation from the HQ limit is driven by the datum on the mode $\psi(2S) \to \chi_{c2}(1P) \gamma$. Averaging the results we obtain
\be
\delta_c^{2S1P}=0.238 \pm 0.012 \,\,{\rm GeV^{-1}}\,\,.
\ee
This allows us to predict
\be
{\cal B}(\eta_c(2S) \to h_c(1P) \gamma)=(0.20\pm 0.03)\%\,\,.
\ee
%%%%%%%%%%%%%%%%%%%%%%
%%%%%%%%%%%
\begin{table}[b]
\centering 
\begin{tabular}{l  c } 
\hline \hline
%\rowcolor{blue!20} 
Decay mode & Branching ratio   \\ 
\hline 
$\chi_{b0}(1P)\rightarrow\gamma \Upsilon(1S) $ & $(1.94 \pm 0.27) \% $ \\ 
%\hline 
$\chi_{b1}(1P)\rightarrow\gamma \Upsilon(1S) $ & $(35.2 \pm 2.0) \% $ \\ 
%\hline 
$\chi_{b2}(1P)\rightarrow\gamma \Upsilon(1S) $ & $(18.0 \pm 1.0) \% $ \\ 
%\hline 
$h_b(1P)\rightarrow\gamma \eta_b(1S) $ & $(52 ^{+6}_{-5}) \% $ \\ 
%\hline 
$\Upsilon(2S)\rightarrow\eta_{b}(1S) \gamma $ & $(5.5 ^{+1.1}_{-0.9}) \times 10^{-4} $  \\
%\hline 
$\Upsilon(2S)\rightarrow \gamma \chi_{b0}(1P) $ & $(3.8 \pm 0.4 ) \% $   \\ 
%\hline 
$\Upsilon(2S)\rightarrow \gamma \chi_{b1} (1P) $ & $(6.9 \pm 0.4) \% $ \\ 
%\hline 
$ \Upsilon(2S)\rightarrow \gamma \chi_{b2}(1P) $ & $(7.15 \pm 0.35) \% $  \\ 
%\hline 
 $\chi_{b0}(2P)\rightarrow\gamma \Upsilon(1S) $ & \, $(3.8 \pm 1.7) \times 10^{-3}$ \, \\ 
%\hline 
$\chi_{b0}(2P)\rightarrow\gamma \Upsilon(2S) $ & $(1.38 \pm 0.30) \% $ \\ 
%\hline 
$\chi_{b1}(2P)\rightarrow\gamma \Upsilon(1S) $ & $(9.9 \pm 1.0 ) \%$  \\
%\hline
$\chi_{b1}(2P)\rightarrow\gamma \Upsilon(2S) $ & $(18.1 \pm 1.9 ) \% $ \\ 
%\hline 
$\chi_{b2}(2P)\rightarrow\gamma \Upsilon(1S) $ & $(6.6 \pm 0.8 ) \%$ \\
%\hline 
$\chi_{b2}(2P)\rightarrow\gamma \Upsilon(2S) $ & $(8.9 \pm 1.2) \% $ \\ 
%\hline
$h_b(2P)\rightarrow\gamma \eta_b(1S)$ & $(22 \pm 5) \% $ \\ 
%\hline 
$h_b(2P)\rightarrow\gamma \eta_b(2S)$ & $(48 \pm 13) \% $ \\ 
\hline \hline

\end{tabular} 
\caption{Radiative branching ratios of $b\bar{b}$ mesons \cite{ParticleDataGroup:2024cfk}. } 
\label{BRbbbar} 
\end{table} 

Let us now consider  radiative transitions of the states in the multiplet $[\chi_{c0}(2P),\,\chi_{c1}(2P),\,\chi_{c2}(2P),\,h_c(2P)]$. If one identifies $\chi_{c1}(3872)$ with $\chi_{c1}(2P)$, the only available datum is the branching ratio ${\cal B}(\chi_{c1}(3872) \to J/\psi \gamma)$ quoted in Table \ref{BRccbar}.
Using this result we determine
\be
\delta_c^{2P1S}=0.018 \pm 0.004 \, \, {\rm GeV}^{-1} \,\,.
\ee
Using this value the prediction follows
\be
{\cal B}(\chi_{c2}(2P) \to J/\psi \gamma)=(3.1 \pm 1.5) \times 10^{-4} \label{chic22P}\,\,.
\ee
This reflects our working hypothesis of identifying $\chi_{c1}(3872)$ with $\chi_{c1}(2P)$: the measurement of this observable as predicted in Eq.~\eqref{chic22P} would  support such an identification.
A prediction could be worked out in the case of decaying $\chi_{c0}(2P)$, but the large experimental error on the full width of this particle would make the result of limited use.
%%%

In the beauty sector the analogous predictions are limited by the incomplete knowledge of the full width of several bottomonia states. 
Therefore, we can quote our results modulo the total width of the decaying particle.
Considering the transitions from the multiplet $[\chi_{b0}(1P),\,\chi_{b1}(1P),\,\chi_{b2}(1P),\,h_b(1P)]$ to the doublet $[\eta_b(1S),\,\Upsilon(1S)]$, we define ${\tilde \delta}_b^{nPmS}(P_b)=\displaystyle\frac{\delta^{nPmS}_b}{[\Gamma_{\rm tot}(P_b)]^{1/2}}$, where $P_b$ generically denotes one of the states in the $nP$ multiplet.  We find
\bea
{\tilde \delta}_b^{1P1S} (\chi_{b0}(1P))&=&(1.79 \pm 0.12)\,\, {\rm GeV}^{-3/2}\nn\\
{\tilde \delta}_b^{1P1S} (\chi_{b1}(1P))&=&(6.77 \pm 0.19) \,\, {\rm GeV}^{-3/2} \nn\\
{\tilde \delta}_b^{1P1S}(\chi_{b2}(1P)) &=&(4.54 \pm 0.13)\,\, {\rm GeV}^{-3/2} \\
{\tilde \delta}_b^{1P1S} (h_b(1P))&=&(6.7 \pm 0.4)\,\, {\rm GeV}^{-3/2} \nn \,\,.
\eea
The same procedure can be applied to the radiative transitions of the states $[\chi_{b0}(2P),\,\chi_{b1}(2P),\,\chi_{b2}(2P),\,h_b(2P)]$ to the two doublets $[\eta_b(1S),\,\Upsilon(1S)]$ and $[\eta_b(2S),\,\Upsilon(2S)]$,
giving
\bea
{\tilde \delta}_b^{2P1S} (\chi_{b0}(2P))&=&(0.31 \pm 0.07) \,\, {\rm GeV}^{-3/2} \nn\\
{\tilde \delta}_b^{2P1S} (\chi_{b1}(2P))&=&(1.51 \pm 0.08)\,\, {\rm GeV}^{-3/2} \nn\\
{\tilde \delta}_b^{2P1S} (\chi_{b2}(2P))&=&(1.20 \pm 0.07) \,\, {\rm GeV}^{-3/2} \\
{\tilde \delta}_b^{2P1S} (h_b(2P))&=&(2.01 \pm 0.23) \,\, {\rm GeV}^{-3/2} \nn
\eea
\noindent and 
\bea
{\tilde \delta}_b^{2P2S} (\chi_{b0}(2P))&=&(3.9 \pm 0.4) \,\, {\rm GeV}^{-3/2} \nn\\
{\tilde \delta}_b^{2P2S} (\chi_{b1}(2P))&=&(12.0 \pm 0.6) \,\, {\rm GeV}^{-3/2} \nn\\
{\tilde \delta}_b^{2P2S}(\chi_{b2}(2P)) &=&(7.8 \pm 0.5) \,\, {\rm GeV}^{-3/2} \\
{\tilde \delta}_b^{2P2S}(h_b(2P)) &=&(16.5 \pm 2.2) \,\, {\rm GeV}^{-3/2} \nn \,\,.
\eea
Defining
\be
R_Q^\delta=\frac{\delta_Q^{2P2S}}{\delta_Q^{2P1S}}
\ee
\noindent
and noticing that  $\displaystyle{\frac{\delta_Q^{2P2S}(P_b)}{\delta_Q^{2P1S}(P_b)}=\frac{{\tilde \delta}_Q^{2P2S}(P_b)}{{\tilde \delta}_Q^{2P1S}(P_b)}}$ since the dependence on the full width cancels out,   we are able to predict
\bea
R_b^\delta(\chi_{b0}(2P))&=&13 \pm 4
\nn \\
R_b^\delta(\chi_{b1}(2P))&=&8 \pm 1
\nn \\
R_b^\delta(\chi_{b2}(2P))&=&7 \pm 1
 \\
R_b^\delta(h_b(2P))&=&8 \pm 2
\nn \,\,.
\eea
In parenthesis we have indicated the decaying particle whose radiative transitions are used to determine the value of $R_b^\delta$. Averaging these results we obtain
\be
R_b^\delta=9.0 \pm 0.7 \label{Rb} \,\,.
\ee
We have stressed that for heavy quarkonia the heavy quark flavor symmetry is not applicable.
On general grounds one expects that violations of HQ symmetries can be quantified  by a suppression factor proportional to the inverse powers of the heavy quark mass. 
Therefore, while we expect $\delta_c^{nPnS} \neq \delta_b^{nPmS}$, it is reasonable to assume that  the parameters $\delta$ obey the scaling rule $\delta_Q \simeq 1/m_Q$, so that $R_b^\delta \simeq R_c^\delta$. This allows us to derive several predictions from the result \eqref{Rb}.

In terms of the ratio $R_c^\delta$ one has
\bea
{\cal R}(\chi_{c1}(3872))&=&\displaystyle\frac{{\cal B}(\chi_{c1}(3872) \to \psi(2S) \gamma)}{{\cal B}(\chi_{c1}(3872) \to J/\psi \gamma)}\nn \\&=&(PS) \left[R_c^\delta (\chi_{c1}(3872))\right]^2 \,\, ,
\eea
where the  constant (PS) is only due to the different  phase space in the two modes.

We have stated in the Introduction that the measurement of ${\cal R}(\chi_{c1}(3872))$
 is of primary interest,  since the various interpretations for this state predict different values for the ratio that should be compared to the updated experimental determination by the LHCb Collaboration \cite{LHCb:2024tpv}.
Adopting the scaling rule stated above we predict
\be
{\cal R}(\chi_{c1}(3872))=1.7 \pm 0.3 \,\,\,. \label{r1}
\ee
 This result on one hand confirms the prediction in \cite{DeFazio:2008xq}, on the other hand shows a very good  agreement with the  LHCb  measurement \cite{LHCb:2024tpv} (last entry in Table \ref{tabRexp}),
 \be
 {\cal R}(\chi_{c1}(3872))_{\rm LHCb}=1.67 \pm 0.21 \pm 0.12 \pm0.04  \,\,\,.
 \label{lhcb}
 \ee
Using the result \eqref{r1} together with the experimental datum for ${\cal B}(\chi_{c1}(3872) \to J/\psi \gamma)$ we obtain
\be
{\cal B}(\chi_{c1}(3872) \to \psi(2S) \gamma)=(1.3 \pm0.5)\times 10^{-2}\,\,.
\ee
The same ratio can be predicted for the other states in the $2P$ multiplet, with the result
\bea
{\cal R}(\chi_{c0}(2P))&=& 1.5 \pm 0.2 \label{r0} \\
{\cal R}(\chi_{c2}(2P))&=& 2.9 \pm 0.4 \,\,.\label{r2}
\eea
Let us stress  that these predictions are correlated, they stem from the assumption that the various particles are indeed the charmonia populating the $2P$ multiplet. They represent a  test of the charmonium option for $\chi_{c1}(3872)$.

We complete the predictions derived from Lagrangian \eqref{lagPS} considering the decays of the beauty states in the $2S$ doublet to the ones in the $1P$ multiplet.
Experimental data for the branching ratios of such processes can be found in Table~\ref{BRbbbar}.
Comparing them with Eqs.~\eqref{deltanPmS2}
we find the results in Table~\ref{valuesdeltab2s1p}. 
\begin{table}[t]
\centering 
\begin{tabular}{c c } 
\hline \hline
%\rowcolor{gray!20} 
Decay mode & $\delta_{b}^{2S1P}$ \\ 
\hline 
$\Upsilon(2S) \rightarrow \chi_{b0}(1P) \gamma $ & $(9.0 \pm 0.6) \times 10^{-2} \, \text{GeV}^{-1}$ \\
%\hline 
$\Upsilon(2S) \rightarrow \chi_{b1}(1P)  \gamma $ & \,$(1.00 \pm 0.05) \times 10^{-1} \, \text{GeV}^{-1}$ \,\\
%\hline 
$\Upsilon(2S) \rightarrow \chi_{b2}(1P)  \gamma $ & $(9.7 \pm 0.4) \times 10^{-2} \, \text{GeV}^{-1}$ \\
%\hline 
%\rowcolor{gray!20} 
\textbf{} & Average \\ 
%\hline 
& $ (9.6 \pm 0.3) \times 10^{-2} \, \text{GeV}^{-1}$\\
\hline \hline
\end{tabular} 
\caption{Results for the coupling $\delta_{b}^{2S1P}$.} 
\label{valuesdeltab2s1p}
\end{table} 
The average value of the coupling  corresponds to
\be
\delta_{b}^{2S1P}=(9.6 \pm 0.3) \times 10^{-2} \, \text{GeV}^{-1} \,\,.
\ee
From this result we  predict  the branching ratio $\mathcal{B}(\eta_{b}(2S) \rightarrow h_b(1P) \gamma)$  for which no data are available:
\be
\mathcal{B}(\eta_{b}(2S) \rightarrow h_b(1P)\gamma)=(2.85\pm0.2)\times 10^{-6}.
\ee

We now consider the transitions governed by the Lagrangian \eqref{lagSS}. Comparing the data in Tables \ref{BRccbar} and \ref{BRbbbar} to the result in \eqref{deltanSmS}
we can  compute the couplings $\delta_c^{nSmS}$ and $\delta_b^{nSmS}$.
We obtain
\begin{itemize}
\item
from the decay $J/\psi \to \eta_c(1S) \gamma$:  
\be
\delta_c^{1S1S} = (4.85 \pm 0.3) \times 10^{-2}\ \text{GeV}^{-1}\,\,;
\ee
\item
from the decay $\psi(2S) \to \eta_c(1S) \gamma$:  
\be
\delta_c^{2S1S} = (3.42 \pm 0.26) \times 10^{-3} \ \text{GeV}^{-1}
\,\,;
\ee
\item
from the decay $\psi(2S) \to \eta_c(2S) \gamma$:  
\be
\delta_c^{2S2S} = (0.66 \pm 0.24) \times 10^{-1} \ \text{GeV}^{-1}\,\,;
\ee
\item
from the decay $\Upsilon(2S) \to \eta_b(1S) \gamma$:  
\be
\delta_b^{2S1S} = (4.50 \pm 0.05) \times 10^{-4} \ \text{GeV}^{-1} \,\,.
\ee
\end{itemize}
We have already observed that the Lagrangian \eqref{lagSS} breaks spin symmetry. Following our previous argument, we expect that the couplings $\delta$ scale as $1/m_Q$. However, when comparing the results in the beauty and charm case, the flavor breaking should also be taken into account, so that we guess the  scaling for the ratios $\delta_b^{nSmS}/\delta_c^{nSmS}\simeq (m_c/m_b)^2$.
Therefore, we expect  $\delta_b^{nSmS}/\delta_c^{nSmS}\simeq (\delta_b^{nPmS}/\delta_c^{nPmS})^2$.
We can test this expectation using our outcome for the $2S \to 1P$ and $2S \to 1S$ transitions.
We find
 \be
\left(\frac{\delta_b^{2S1P}}{\delta_c^{2S1P}}\right)^2/ \left(\frac{\delta_b^{2S1S}}{\delta_c^{2S1S}}\right)=1.25 \pm 0.2 \,.
 \ee
 The result supports our assumption about the scaling of the parameters $\delta$ and, consequently, the hypothesis $R_b^\delta=R_c^\delta$ that has led to the prediction \eqref{r1}.
 
\section{Comparison with other approaches and conclusions}
The discriminating power of the result for the ratio ${\cal R}(\chi_{c1}(3872))$ has first been put forward in \cite{Swanson:2004pp}. 
In multiquark scenarios,
both in the compact tetraquark and in the molecular cases, light quarks are part of the structure of $\chi_{c1}(3872)$, which implies that other mechanisms are possible for radiative transitions with respect to the case in which only a $c{\bar c}$ component is present.
In the molecular scenario, radiative decays could originate from vector meson dominance mechanism or from light quark annihilation, generally producing results ${\cal R}(\chi_{c1}(3872))\ll 1$  \cite{Dong:2009uf,Guo:2014taa,Cincioglu:2016fkm,Takeuchi:2016hat,Lebed:2022vks,Grinstein:2024rcu}. In   a potential model framework  \cite{Grinstein:2024rcu}, modifying the  extension of the wave functions  of the involved hadrons in the molecular and in the compact tetraquark case  produces ${\cal R}(\chi_{c1}(3872))\simeq {\cal O}(1)$ for the compact tetraquark.\footnote{  A  recent refinement of the analysis in \cite{Grinstein:2024rcu}   gives ${\cal R}(\chi_{c1}(3872))=1.4 \pm 0.3$ \cite{Germani:2025qhg}.}
The possibility of a mixing between exotic and ordinary charmonium components has been considered in the framework of the Born-Oppenheimer effective field theory, finding that the charmonium component is essential to reproduce the measured value of the ratio ${\cal R}(\chi_{c1}(3872))$  \cite{Brambilla:2024thx}.

In the conventional charmonium interpretation, various results have been obtained in different variants of the potential models, that are largely dependent on the details of the model, in particular on the wave functions of the charmonia involved in the decays \cite{Barnes:2003vb,Barnes:2005pb,Li:2009zu,Badalian:2012jz,Ferretti:2014xqa,Badalian:2015dha,Deng:2016stx}. 

 In contrast to such approaches, the present study has been carried out in an effective Lagrangian framework that exploits the heavy quark spin symmetry  arising in QCD in the large heavy quark mass limit, hence it is  model independent to a large extent. Moreover,  there are no adjustable parameters. 
In this framework we have worked out the consequences of the identification of $\chi_{c1}(3872)$ with $\chi_{c1}(2P)$. 
The result for the ratio ${\cal R}(\chi_{c1}(3872))$ 
confirms the outcome obtained in \cite{DeFazio:2008xq} and favorably compares with the latest experimental  determination. 
This demonstrates the reliability of the method, since it is stable against small variations of the input experimental values.
It should be stressed that the agreement of our result with data has improved with respect to  the 2008 analysis, since the recent LHCb measurement is closer to our prediction than the BABAR Collaboration result available at that time. 

We have worked out other predictions involving the states in the $2P$ $c{\bar c}$ fourplet that will be compared to data when  available. The whole pattern of predictions will serve as an indirect support to the identification of $\chi_{c1}(3872)$ with one of the members of that multiplet. 

Other issues related to $\chi_{c1}(3872)$ need to be answered before reaching a consensus on its identification with $\chi_{c1}(2P)$. However, if the answers would not be convincing and the identification would not be accepted, a question would remain: where is $\chi_{c1}(2P)$?
%\vspace*{0.5cm}

\acknowledgements
We thank Marco Pappagallo for discussions.
This study has been  carried out within the INFN project (Iniziativa Specifica) SPIF.
The research has been partly funded by the European Union – Next Generation EU through the research Grant No. P2022Z4P4B “SOPHYA - Sustainable Optimised PHYsics Algorithms: fundamental physics to build an advanced society" under the program PRIN 2022 PNRR of the Italian Ministero dell’Universit\'a e Ricerca (MUR).

\bibliographystyle{apsrev4-1}
\bibliography{refFGP}
\end{document}